\documentclass[a4paper,11pt]{article}
\usepackage{jinstpub}
\usepackage{lineno}
\usepackage{booktabs}
\usepackage{graphicx}
\usepackage{float}
\usepackage{subfigure}
\usepackage{multirow}

\title{Studying single-electron traps in newly fabricated Skipper-CCDs for the Oscura experiment using the pocket-pumping technique}

\author[a,b,c]{S. E. Perez}
\author[a]{B. A. Cervantes-Vergara}
\author[a]{J. Estrada}
\author[d]{S. Holland}
\author[b,c]{D. Rodrigues}
\author[a]{J. Tiffenberg}

\affiliation[a]{Fermi National Accelerator Laboratory, IL, USA}
\affiliation[b]{Universidad de Buenos Aires, Facultad de Ciencias Exactas y Naturales, Departamento de Fisica. Buenos Aires, Argentina.}
\affiliation[c]{CONICET - Universidad de Buenos Aires, Instituto de Fisica de Buenos Aires (IFIBA). Buenos Aires, Argentina}
\affiliation[d]{Lawrence Berkeley National Laboratory, CA, USA}

\emailAdd{santiep.137@gmail.com, bcervant@fnal.gov}

\abstract{Understanding and characterizing very low-energy ($\sim$eV) background sources is a must in rare-event searches. Oscura, an experiment aiming to probe electron recoils from sub-GeV dark matter using a 10-kg skipper-CCD detector, has recently fabricated its first two batches of sensors. In this work, we present the characterization of defects/contaminants identified in the buried-channel region of these newly fabricated skipper-CCDs. These defects/contaminants produce deferred charge from trap emission in the images next to particle tracks, which can be spatially resolved due to the sub-electron resolution achieved with these sensors. Using the trap-pumping technique, we measured the energy and cross section associated to these traps in three Oscura prototype sensors from different fabrication batches which underwent different gettering methods during fabrication. Results suggest that the type of defects/contaminants is more closely linked to the fabrication batch rather than to the gettering method used. The exposure-dependent single-electron rate (SER) of one of these sensors was measured $\sim$100~m underground, yielding $(1.8\pm 0.3)\times10^{-3}~e^-$/pix/day at 131K. The impact of the identified traps on the measured exposure-dependent SER is evaluated via a Monte Carlo simulation. Results suggest that the exposure-dependent SER of Oscura prototype sensors would be lower in lower background environments as expected.}


\begin{document}
\maketitle
\flushbottom

\section{Introduction} \label{sec:intro}
Since their invention in 1969, Charge-Coupled Devices (CCDs) have been widely adopted in space and ground based astronomical surveys. They possess appealing characteristics such as a spatial resolution as low as a few $\mu$m, low readout noise and a low dark-count rate. Recently, the skipper-CCD~\cite{Janesick1990, Tiffenberg2017}, with enhanced sensitivity to low-energy signals, has become one of the most promising technologies for Dark Matter (DM) and rare-event searches. In these applications, the discovery potential is highly constrained by the one-electron background rate~\cite{Rouven2016}.

Many background sources of Single-Electron Events (SEEs) in skipper-CCD detectors have been identified and characterized, including temperature fluctuations, radiative processes from external radiation interactions, low-energy photons from the amplifiers and clock-induced charge~\cite{Janesick2001, sensei2022, Rouven2024}. However, we have recently identified another source of SEEs in the newly fabricated skipper-CCDs for Oscura~\cite{OscuraSensors2023}, a multi-kilogram experiment aiming to probe electron recoils from sub-GeV DM. We associate this source to defects/contaminants within the CCD buried-channel that create single-electron traps with release times comparable to consecutive pixel-readout time, causing a ``tail'' of deferred single-electron depositions after particle tracks. In some cases, this charge can spread within the image, leading to an apparent increase in the exposure-dependent single-electron rate (SER), which might be mistaken for the sensor's intrinsic dark current (DC). 

In this work, we perform the established trap pumping technique~\cite{Blouke1988, Janesick2001, Hall2014, Bilgi2019} to three different Oscura prototype sensors to characterize their buried-channel single-electron traps. We measure the energy and cross section of the main trap species found in the prototype sensors and verify the effect of deferred charge from trap emission on the measured exposure-dependent SER through a Monte Carlo simulation.

\section{Charge trapping characterization in CCDs} \label{sec:trapcharacterization}
\subsection{Shockley-Read-Hall theory}~\label{sec:shockley}
Traps associated to intermediate energy levels within the Si bandgap are usually modeled using the Shockley-Read-Hall model for carrier generation and recombination~\cite{Shockley1952}. The traps lying within the CCD charge-transfer region could capture charge carriers from charge packets as they are transferred through the device, and release them at a later time. The probability of a trap to capture (c) or emit (e) one charge carrier within the time interval [$t_1, t_2$] is given by
\begin{equation} \label{eq:probtraps}
P_{c,e}=e^{-t_{1}/\tau_{c,e}}-e^{-t_{2}/\tau_{c,e}}\,.
\end{equation}
with $\tau_{c,e}$ the characteristic capture (emission) time constant, which can be expressed as
\begin{equation} \label{eq:tautraps}
\tau_c=\frac{1}{\sigma v_{th} n} \qquad {\rm and} \qquad \tau_e=\frac{1}{\sigma v_{th} N_c}e^{\frac{E_t}{k_BT}}\,.
\end{equation}
Here, $T$ is temperature [K], $E_t$ is the trap energy level [eV], $\sigma$ is the trap cross section [cm$^2$], $v_{th}$ is the charge-carrier's thermal velocity [cm/s], $n$ is the charge-carrier concentration in the vicinity of the trap [cm$^{-3}$], and $N_c$ is the effective density of states in the conduction band [cm$^{-3}$]. $v_{th}$ and $N_c$ depend on $T$ and on the charge-carrier's effective mass for conductivity $m_{\rm cond}$ and for density of states $m_{\rm dens}$ calculations as
\begin{equation} \label{eq:DOS}
v_{th}=\sqrt{3k_BT/m_{\rm cond}}\qquad {\rm and} \qquad N_c=2\left[2\pi (m_{\rm dens}) \frac{k_BT}{h^2}\right]^{3/2}\,.
\end{equation}

In p-channel CCDs, charge carriers are holes for which $m^h_{\rm cond}\simeq 0.41 m_e$ and $m^h_{\rm dens}\simeq 0.94 m_e$ between 100K and 200K~\cite{green1990intrinsic}, with $m_e$ the free electron rest mass.

\subsection{Pocket-pumping technique}\label{sec:PocketPumping}
\begin{figure}[!ht]
    \centering
    \includegraphics[width=0.9\linewidth]{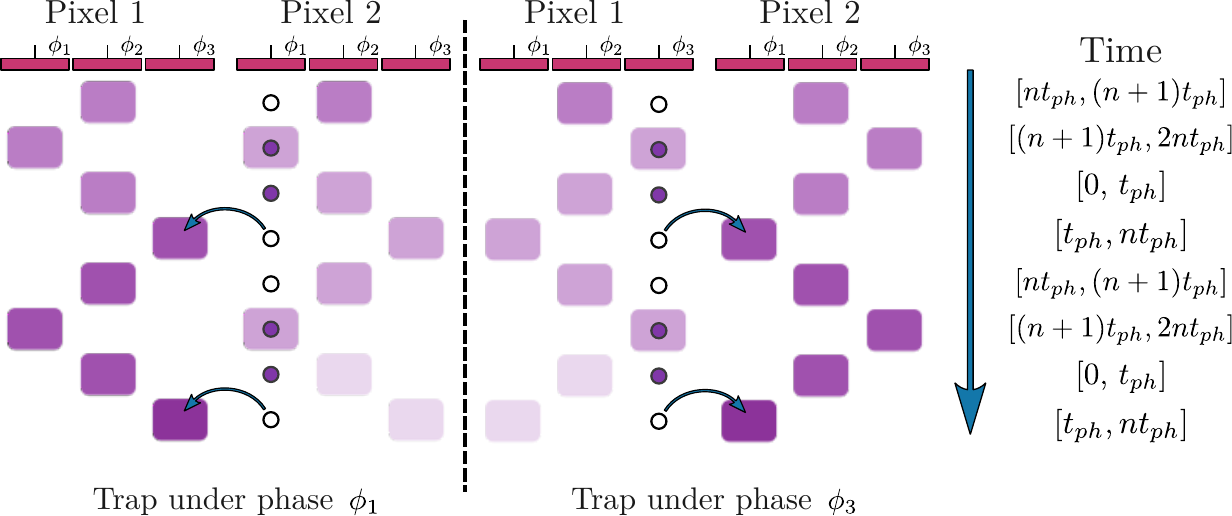}
    \caption{Sub-sequence of states of a three-phase pumping sequence to identify traps under phases $\phi_1$ and $\phi_3$. Here, closed white (purple) circles represent an empty (filled) trap, purple rectangles represent charge packets, with their intensity associated to their amount of charge, and arrows represent trap emission. The figure on the left (right) shows a trap under phase $\phi_1$ ($\phi_3$) that is being filled multiple times. In this sequence, $\phi_2$ would be the ``transient'' phase.}
    \label{fig:trappumpscheme}
\end{figure}

The technique of pocket pumping~\cite{Blouke1988, Janesick2001, Hall2014, Bilgi2019} has proved to be a powerful tool to spatially localize and measure the characteristic parameters of charge traps lying within the CCD charge-transfer region. This method consists of filling the traps by ``uniformly'' illuminating the active area of the CCD and allowing them to emit the trapped charge into their neighbor pixel multiple times. This is done by repeatedly moving the charge back and forth, between pixel phases, creating ``dipole'' signals relative to the flat background. The method is illustrated in Fig.~\ref{fig:trappumpscheme}. The sequence of states in this figure is useful to detect traps located below phases $\phi_1$ and $\phi_3$ in a three-phase device.

Within the trap pumping sequence, charge capture occurs during the state in which charge remains under the phase with the trap. Assuming a 100\% probability of capture, the emission clock starts running just after charge is moved from the phase with the trap, going through the ``transient'' phase(s). The state in which trap emission takes place corresponds to the one in which charge from the adjacent pixel to the pixel with the trap lies in the adjacent phase to the phase with the trap. The effective time spent in this state can be considered to be a multiple integer of $t_{ph}$, which is the time spent under the ``transient'' phase(s). Particularly, for the pumping sequence shown in Fig.~\ref{fig:trappumpscheme}, the time interval spent in this state is $\left[t_{ph},~nt_{ph}\right]$ and the probability of emission is given by Eq.~\eqref{eq:probtraps} evaluated within this time interval. The emission clock resets after each pumping cycle, when charge passes again through the trap.

After completing a given number of pumping cycles $N_{\rm pumps}$, the intensity of the dipole signal, composed of a bright (b) and a dark (d) pixel with $S_b$ and $S_d$ charge carriers, respectively, can be expressed as
\begin{equation} \label{eq:dipintens}
    I_{\rm dip}=\frac{1}{2}|S_b-S_d|=N_{\text{pumps}}D_tP_cP_e\,,
\end{equation}
where $D_t$ is the trap depth. Here, the probability of the trap to capture a charge carrier $P_c$ has been incorporated as a linear scaling factor~\cite{Hall2014}.
The time spent in the state in which trap emission takes place can be optimized to minimize the total time of the pumping sequence to achieve the maximum dipole intensity $\mathcal{T}\vert_{I^{\rm max}_{\rm dip}}$. In the case of the three-phase pumping sequence illustrated in Fig.~\ref{fig:trappumpscheme}, $\mathcal{T}=2nt_{ph}N_{\rm pumps}$. From Eq.~\ref{eq:dipintens} and assuming $I_{\rm dip}\propto P_e$, the maximum intensity $I^{\rm max}_{\rm dip}$ occurs at $t_{ph}\vert_{I^{\rm max}_{\rm dip}}=\tau_e \ln{n}/(n-1)$; note that for higher values of $n$, $I^{\rm max}_{\rm dip}$ happens at lower $t_{ph}$. Given $I^{\rm max}_{\rm dip}$, $N_{\rm pumps}\vert_{I^{\rm max}_{\rm dip}}\propto n^{n/(n-1)}/(n-1)$. Hence, the minimum of $\mathcal{T}\vert_{I^{\rm max}_{\rm dip}}$ is achieved when $n=8$~\cite{Bilgi2019}.
Using the optimization described above, for a given t$_{ph}$ Eq.~\ref{eq:dipintens} takes the form
\begin{equation} \label{eq:dipintens_replaced}
    I_{\rm dip}=N_{\text{pumps}}D_tP_c\big(e^{-\frac{t_{ph}}{\tau_e}}-e^{-8\frac{t_{ph}}{\tau_e}}\big)\,.
\end{equation}

By fitting $I_{\rm dip}$ as a function of $t_{ph}$, the emission constant of an individual trap can be extracted. Furthermore, if data is taken at different temperatures, from the fit of $\tau_e(T)$, given by Eq.~\ref{eq:tautraps}, the energy level and cross section of the trap can be obtained.

\subsection{Effects of charge traps in electron-counting CCDs} \label{sec:trapsignsinCCDs}
Typical images from CCDs used for DM and rare-event searches are dark exposures containing tracks of different particles. In a sensor containing traps within the sensor charge-transfer region, depending on the ratio of the traps characteristic emission time and the readout time between two consecutive pixels $t_{pix}$, trapped charge from these tracks can be emitted: 1) within the pixels of the event, when $\tau_e/t_{pix} \ll 1$; 2) in a highly localized region in the readout direction next to the event, when $\tau_e/t_{pix} \simeq 1$; or 3) after several pixels, when $\tau_e/t_{pix} \gg 1$. Because of the dependence of $\tau_e$ with $T$, i.e. Eq.~\ref{eq:tautraps}, the ``tail'' of deferred charge from trap emission next to an event is expected to span more pixels at lower temperatures.

\begin{figure}[!ht]
    \centering
    \includegraphics[width=0.8\linewidth]{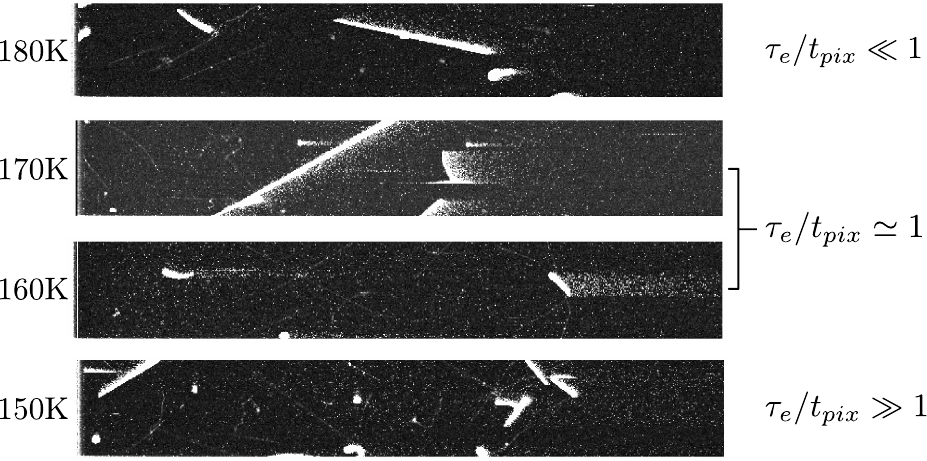}
    \caption{Dark exposure images at different temperatures from the same skipper-CCD with traps within the sensor's charge-transfer region. Tracks from different particles can be seen in the sensor's active area followed by a ``tail'' of deferred charge from trap emission. The sub-electron noise of the image (0.2~$e^-$ with 225~samples/pix) allows to identify deferred single-electron depositions.}
    \label{fig:tail_comp}
\end{figure}

Skipper-CCDs provide a unique tool to resolve unequivocally the spatial distribution of depositions coming from emissions of single-electron traps, due to their sub-electron resolution. This is evident in Fig.~\ref{fig:tail_comp}, where dark exposure images at different temperatures from a skipper-CCD with traps within the sensor charge-transfer region are shown. As these images were taken with multiple samples per pixel, achieving sub-electron noise levels, the spatial distribution of the deferred charge next to particle tracks is resolved.

The identification and subsequent masking of pixels with deferred charge from trap emission is trivial when $\tau_e/t_{pix}\leq 1$ as the deferred charge remains near the main event. However, deferred charge from traps with $\tau_e/t_{pix}\gg1$ cannot be easily identified because of the spatial separation of the deferred charge from the original pixel, and taking a conservative masking approach could lead to a significant loss in exposure. To minimize the span of the deferred charge, $\tau_e$ can be decreased by going to higher temperatures and/or $t_{pix}$ can be increased. With skipper-CCDs the latter can be done by increasing the number of samples per pixel. However, these approaches lead to a background increase from other temperature and/or exposure-dependent sources, which is not desirable in some cases.

\section{Pocket-pumping measurements on Oscura skipper-CCDs}
\subsection{Oscura skipper-CCDs}
The newly fabricated skipper-CCDs for Oscura are 1.35~MPix p-channel CCDs with $15~\mu\textrm{m}\times15~\mu\textrm{m}$ three-phase pixels and a thickness of standard 200-mm silicon wafers (725~$\mu$m)~\cite{OscuraSensors2023}. During the Oscura R\&D phase, two batches of sensors were fabricated using two different extrinsic gettering techniques\footnote{Gettering techniques, implemented during CCD fabrication, create trapping sites for mobile impurities to be drawn away from the active regions of the device. Extrinsic gettering processes create these sites on the back side of the wafer.}~\cite{HOLLAND1989}. All wafers from the first batch and one half of the wafers from the second batch underwent a P ion-implantation induced gettering~\cite{DALLABETTA1997}. The second half of wafers from the second batch underwent a POCl$_3$ induced gettering~\cite{DALLABETTA1997}. In this work, we characterize single-electron traps from three different Oscura prototype sensors, labeled A, B and C in Table~\ref{tab:sensors}, from the two fabricated batches and gettering processes.
\begin{table}[H]
\centering
\begin{tabular}{@{}ccc@{}}
\toprule
Prototype   & Gettering type    & Batch  \\ \midrule
A           & Ion implantation  & First  \\
B           & Ion implantation  & Second \\
C           & POCl$_3$          & Second \\ \bottomrule
\end{tabular}%
\caption{Fabrication details of the Oscura skipper-CCDs used for single-electron traps characterization.}\label{tab:sensors}
\end{table}

\subsection{Data taking}~\label{sec:Datataking}
\begin{figure}[!ht]
    \centering

    \subfigure[$t_{ph}=66~\mu$s]{\includegraphics[width=0.4\textwidth]{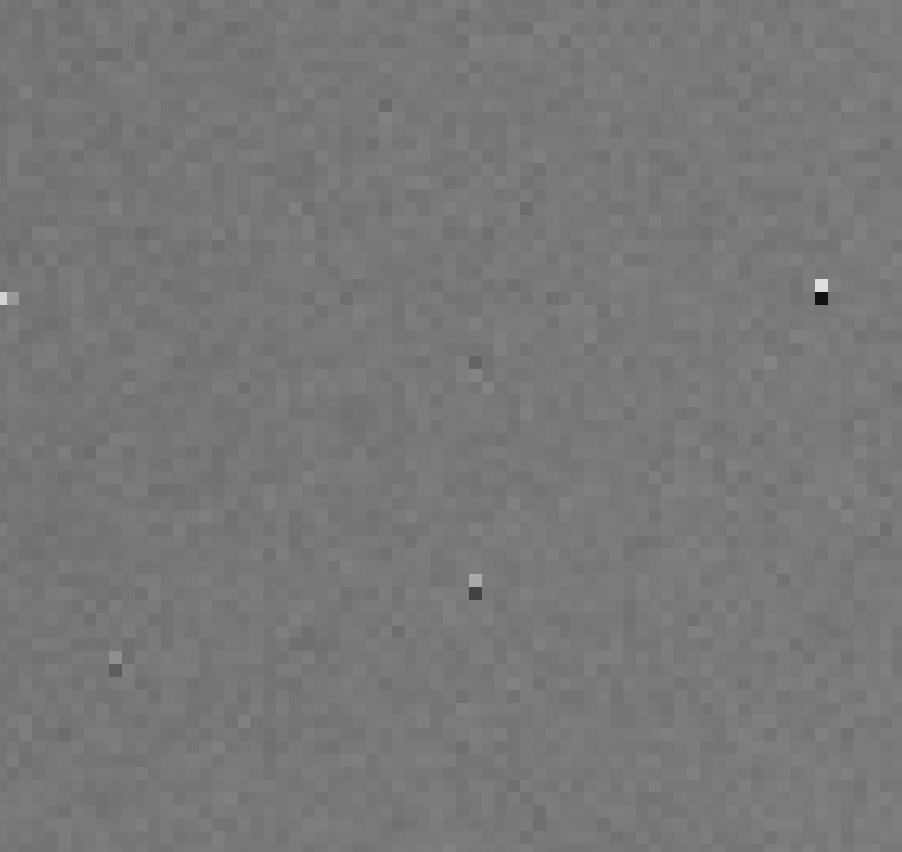}}
    \hspace{1.5cm}
    \subfigure[$t_{ph}=66$~ms]{\includegraphics[width=0.4\textwidth]{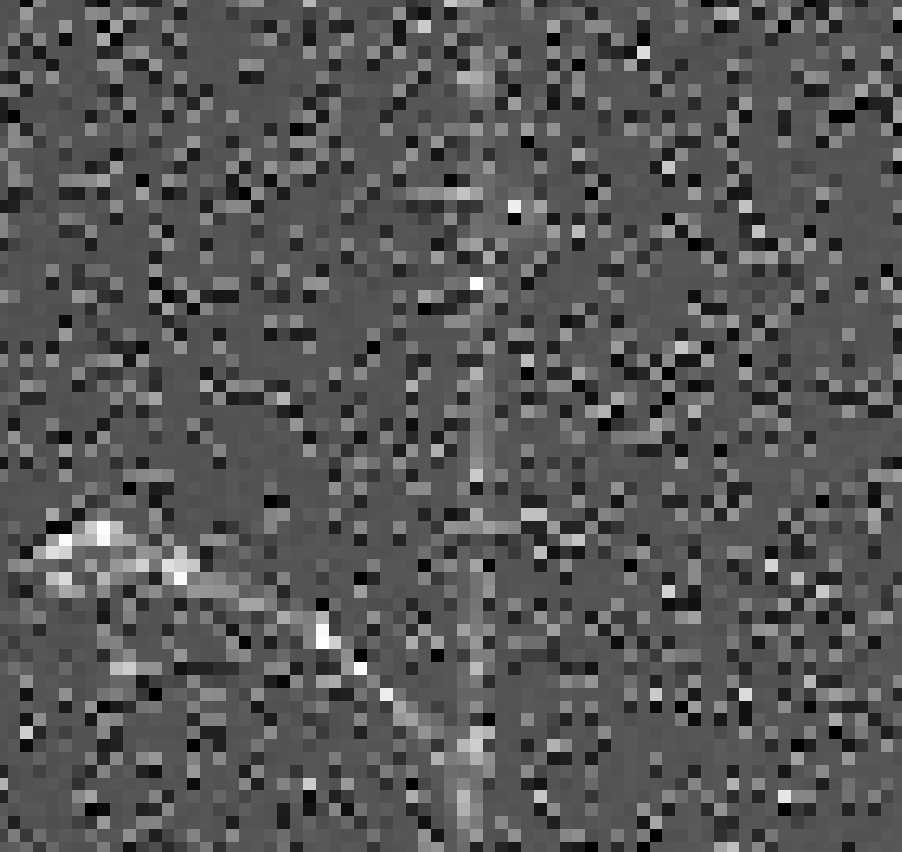}}
    \caption{Left (Right): Section of an image from Oscura prototype sensor A after performing pocket pumping at 150K, for $t_{ph}=66~\mu$s (ms). Dipoles corresponding to charge traps under pixel phases $\phi_1$ and $\phi_3$ are present in both images, but a higher density is evident in the right one.}
    \label{fig:qpump}
\end{figure}

We use the pocket-pumping technique discussed in Section ~\ref{sec:PocketPumping} to localize and characterize traps in the Oscura prototype skipper-CCDs. First, using a violet LED externally controlled by an Arduino Nano, we illuminate the active area of the sensors, which is loosely covered with a Cu plate to increase uniformity in the illumination profile. The median charge per pixel after illumination lies between 1500~$e^-$ and 2000~$e^-$. Then, we perform a pocket-pumping sequence to probe traps below pixel phases $\phi_1$ and $\phi_3$, such as the one illustrated in Fig.~\ref{fig:trappumpscheme}, including the $\mathcal{T}\vert_{I_{\rm max}}$ minimization discussed in Section~\ref{sec:PocketPumping}. We collected images with $N_{\rm pumps}\simeq3000$, varying $t_{ph}$ from 6.6~$\mu$s to 1.3~s and $T$ from 150K to 190K. Fig.~\ref{fig:qpump} shows a section of the images from the pocket-pumping measurements of prototype sensor A at 150K, for two different $t_{ph}$. The right image in this figure reveals a higher density of traps with $\tau_e\sim\mathcal{O}$(ms). We found a uniform spatial distribution of traps through the whole active area of the sensors.

\subsection{Analysis and results}

\begin{figure}[!ht]
    \centering

    \subfigure[]
    {\includegraphics[width=0.5\textwidth]{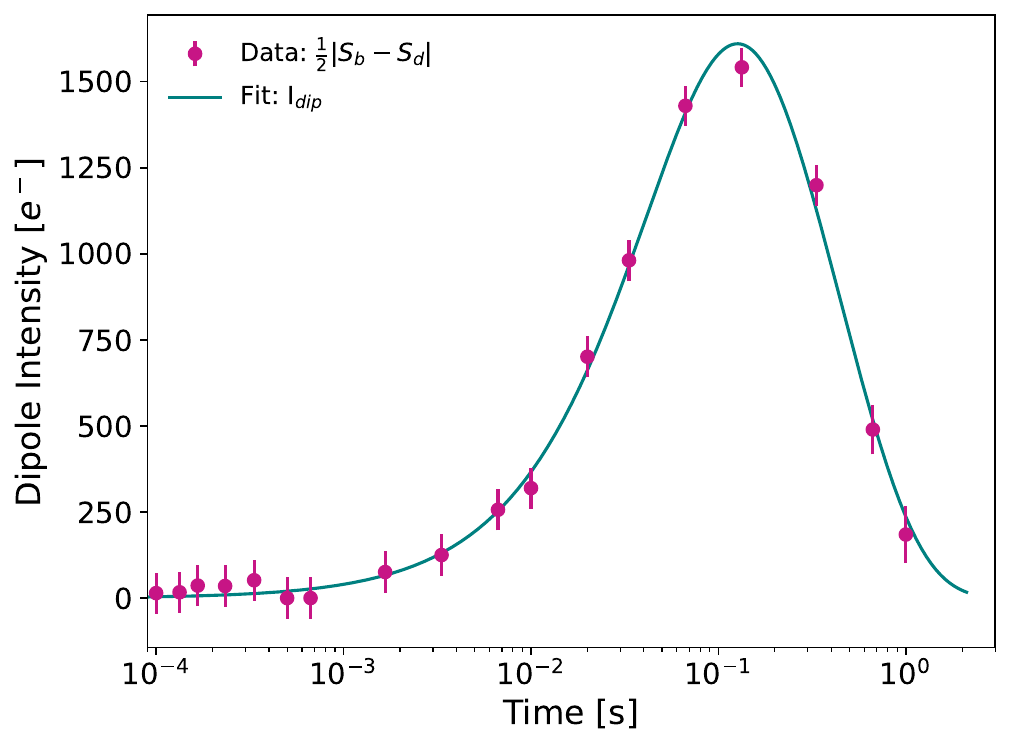}}
    \hspace{1.cm}
    \subfigure[]
    {\includegraphics[width=0.35\textwidth]{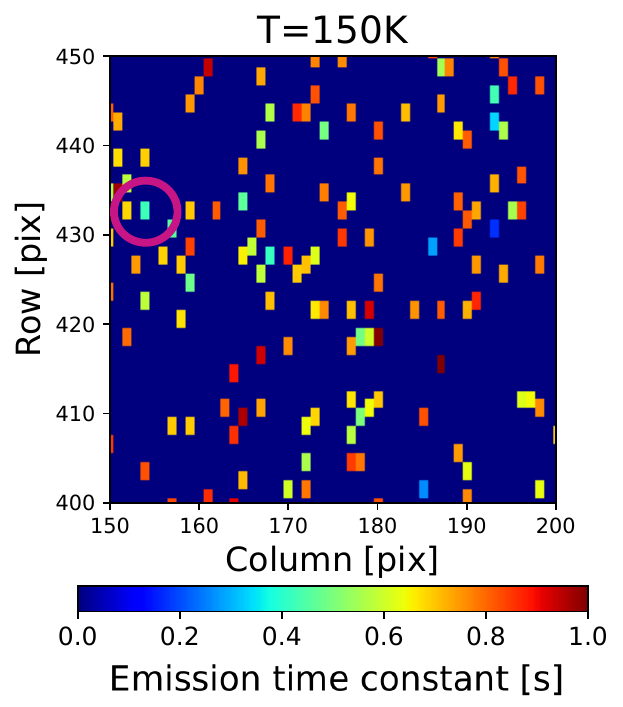}}
    \caption{ Left: Measurements of the dipole intensity versus $t_{ph}$ of a detected trap and their fit with Eq.~\ref{eq:dipintens_replaced}, which leads to $\tau_e=(0.42\pm 0.01)$~s. The errors on the measurements are dominated by Poissonian fluctuations on the pixels' charge. This particular dipole corresponds to the highlighted dipole in the trap map on the right. Right: Map showing the position and the emission-time constant of each detected trap within a region of the active area of the Oscura prototype sensor A at 150K.}
    \label{fig:dipoles_curve}
\end{figure}

With the most efficient dipole-detection algorithm discussed in Appendix~\ref{sec:dipdetection}, we identify dipoles and track their position in each of the images. Using sets of images from the same sensor acquired at a fixed temperature, we compute $I_{\rm dip}$ as a function of $t_{ph}$ for each dipole found. We fit this curve with the function given by Eq.~\ref{eq:dipintens_replaced} and extract the trap emission-time constant $\tau_e$ associated to that dipole. As a quality selection criteria to the dipole intensities, we require a coefficient of determination greater than 0.7 and a relative error on $\tau_e$ below 50\%. With this criteria we reject between 2\% to 20\% of dipoles, depending on the dataset. From now on, we will refer as ``detected traps'' to those probed below pixel phases $\phi_1$ and $\phi_3$ that were found with the trap-detection algorithm, were not rejected by the selection criteria, and are not overlapped dipoles. Figure~\ref{fig:dipoles_curve} (left) shows the intensity as a function of $t_{ph}$ of a detected trap fitted by Eq.~\ref{eq:dipintens_replaced}. For each set of images from the same sensor at a given temperature, we build a trap map with the position and the emission-time constant of each detected trap. One of these maps, from a 50 pix $\times$ 50 pix region of the active area of Oscura prototype sensor A at 150K, is shown in Fig.~\ref{fig:dipoles_curve} (right).

\begin{figure}[!ht]
    \centering
    \subfigure[]{\includegraphics[width=0.49\textwidth]{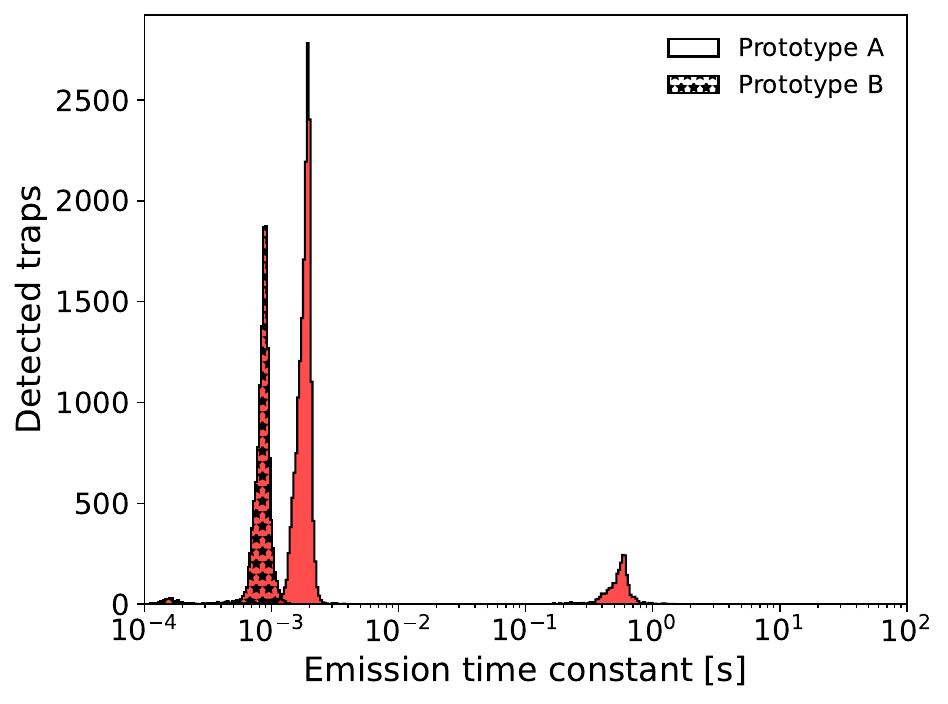}}
    \subfigure[]{\includegraphics[width=0.49\textwidth]{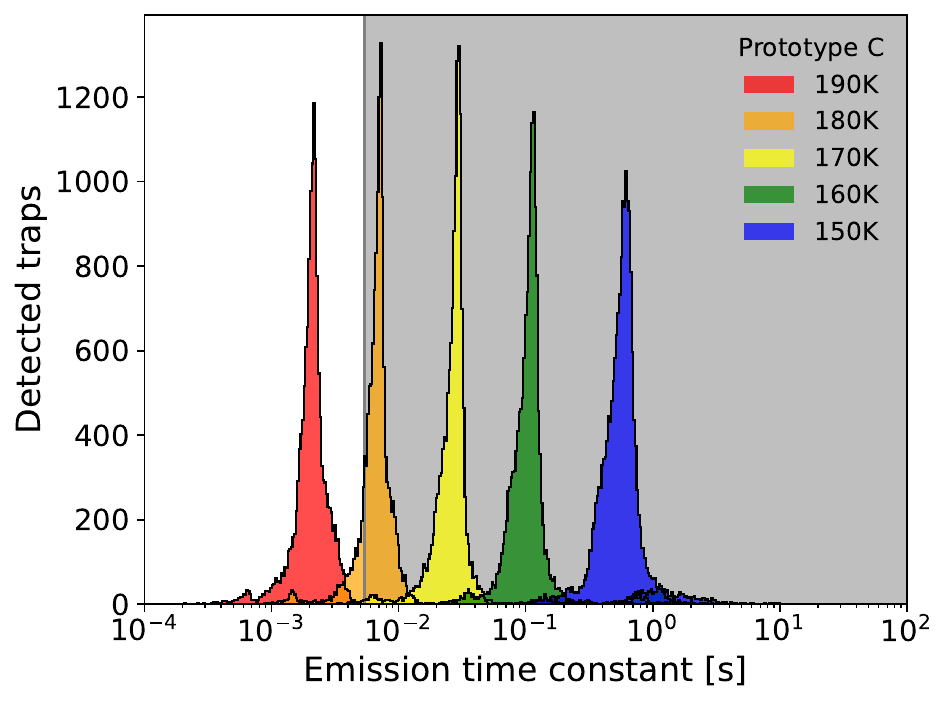}}
    \caption{Left: Distributions of $\tau_e$ at 190K of the detected traps for the Oscura prototype sensors A and B, with the ion-implantation gettering. Right: Distributions of $\tau_e$ at different temperatures of the detected traps for the Oscura prototype sensor C, with the POCl$_3$ gettering. The gray area indicates the $\tau_e$ that are comparable to the expected pixel/image readout time for Oscura, corresponding to the traps that would leave a ``tail'' of deferred charge in the images.}\label{fig:taudist}
\end{figure}

The histograms in Fig.~\ref{fig:taudist} (left) show the $\tau_e$ distributions at 190K of the detected traps for the Oscura prototype sensors A and B, which are from different fabrication batches but underwent the same gettering process (ion implantation). The histograms in Fig.~\ref{fig:taudist} (right) show the $\tau_e$ distributions at different temperatures of the detected traps for the Oscura prototype sensor C, with the POCl$_3$ gettering.

In all the $\tau_e$ distributions in Fig.~\ref{fig:taudist}, a primary peak can be seen, which is associated to the largest population of traps within the sensors' buried-channel region. Also, in 
Fig.~\ref{fig:taudist} (right) the peaks in the distributions move towards higher values of $\tau_e$ at lower temperatures, which is expected from the dependence of $\tau_e$ with $T$, i.e. Eq.~\ref{eq:tautraps}. Comparing the $\tau_e$ distributions from prototype sensors A and B in Fig.~\ref{fig:taudist} (left), both with the ion-implantation gettering, we see a larger population of traps with $\tau_e>0.1$s for $T>170$K in the distributions from prototype sensor A, forming a secondary peak. We associate the presence of this peak to the fabrication batch as none of the distributions from sensors from the 2nd batch, i.e. B and C, show a significant trap population at those $\tau_e$.

\begin{figure}[!ht]
    \centering
    \subfigure[]{\includegraphics[width=0.45\textwidth]{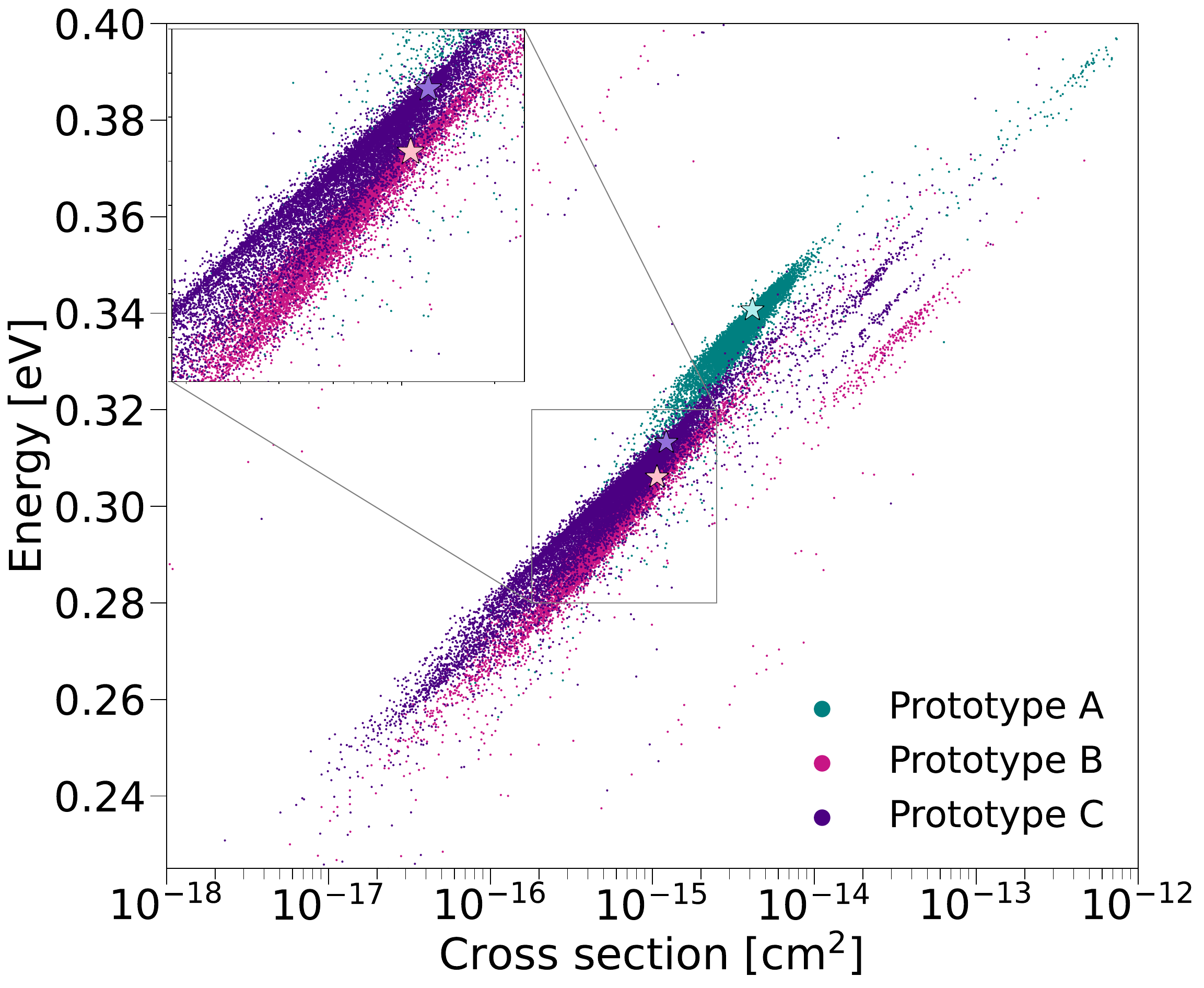}}\hfill
    \subfigure[]{\includegraphics[width=0.49\textwidth]{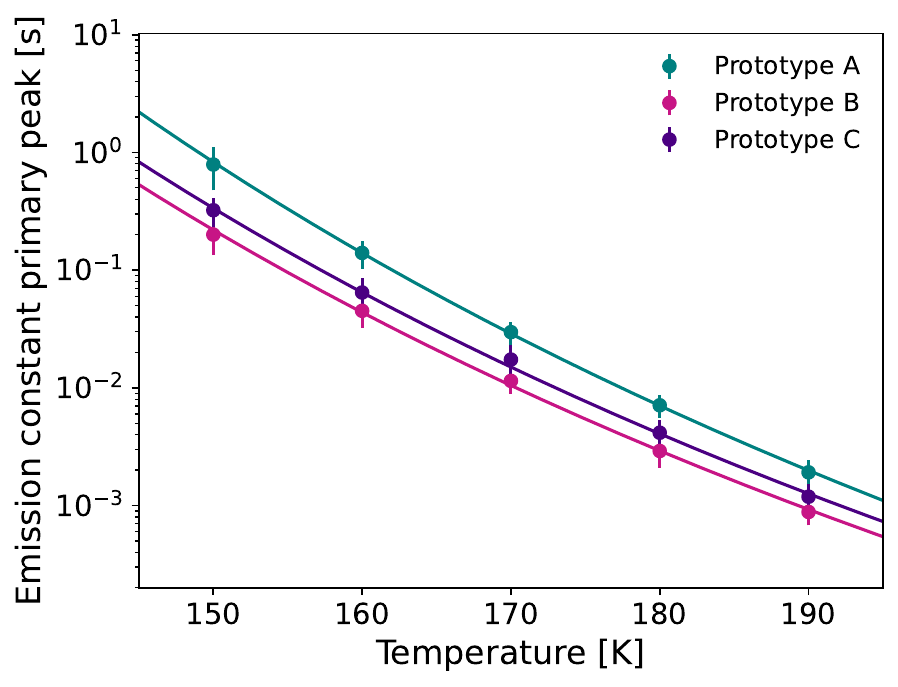}}
    \caption{Left: Scatter plot showing the energy and cross section of each detected trap in the Oscura prototype sensor A (green), B (pink) and C (purple). The values obtained from the fit on the right are plotted as stars. Right: Dependence on $T$ of the primary peaks of the $\tau_e$ distributions in Fig.~\ref{fig:taudist} and their fit with Eq.~\ref{eq:tautraps}. }\label{fig:taufitandcomp}
\end{figure}

\begin{figure}[!ht]
    \centering
    \subfigure[]{\includegraphics[width=0.49\textwidth]{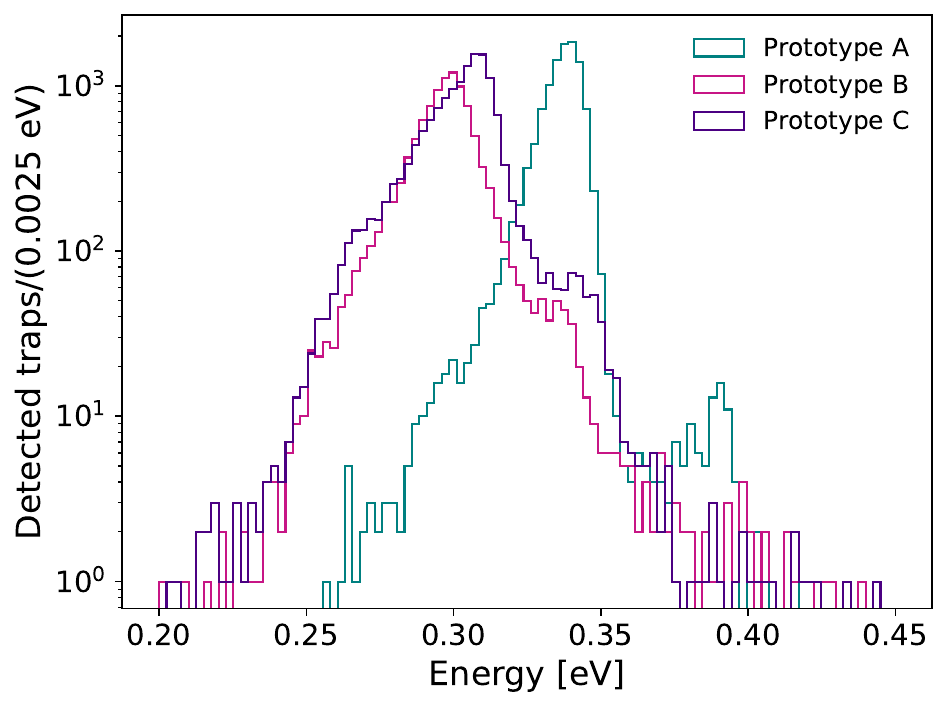}}
    \subfigure[]{\includegraphics[width=0.49\textwidth]{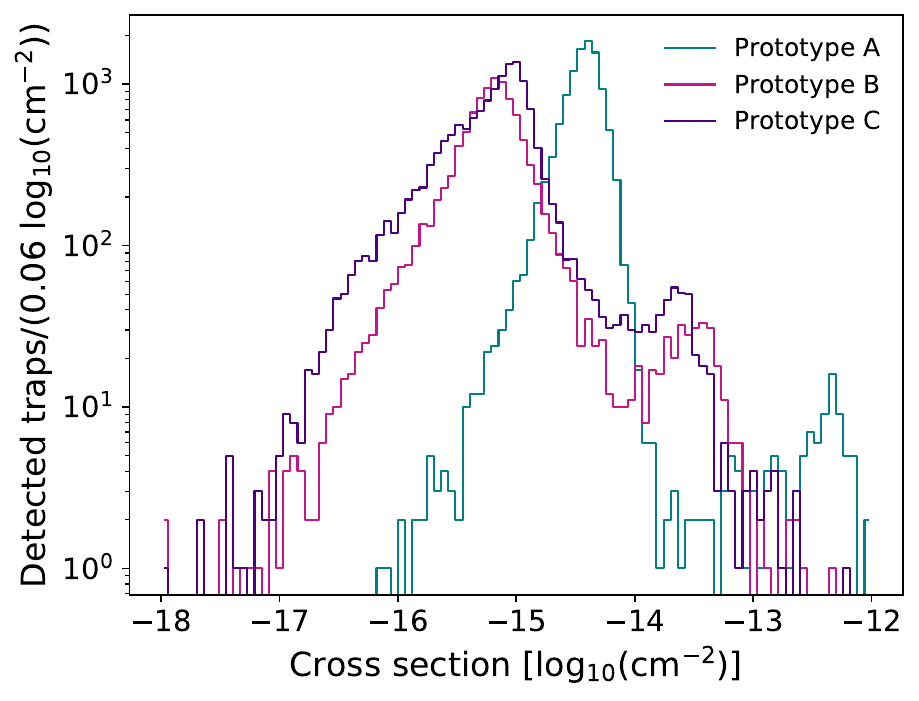}}
    \caption{Distributions of the energy $E_t$ (left) and cross section $\sigma$ (right) of the detected traps in Oscura prototype sensor A (green), B (pink) and C (purple).}\label{fig:trapparams}
\end{figure}

For each detected trap, we plot $\tau_e$ as a function of $T$ and fit it with the function in Eq.~\ref{eq:tautraps}. We perform a chi-squared test on the fits and rejected those with a p-value below 0.05. From each of those fits, we extract the energy $E_t$ and cross section $\sigma$ associated to each trap, shown as dots in the scatter plot in Fig.~\ref{fig:taufitandcomp} (left). The distributions of these variables of the detected traps in each of the Oscura prototype sensors are shown in Fig.~\ref{fig:trapparams}. The maximum value of each of these histograms and its associated error, computed as the full width at half maximum, is shown in Table~\ref{tab:trapparam} [Hist. max.].

Moreover, from the $\tau_e$ distributions at different temperatures associated to each sensor, i.e. histograms in Fig.~\ref{fig:taudist}, we plot the emission-time constants associated to the primary peaks $\tau^{\rm peak}_{e}$ against $T$, with an error given by its full width at half maximum, as shown in Fig.~\ref{fig:taufitandcomp} (right). We fit the data points with the function in Eq.~\ref{eq:tautraps} and extract from it the energy $E_t$ and cross section $\sigma$ associated to the largest population of traps. The value of these variables and its associated error are shown in Table~\ref{tab:trapparam} [$\tau^{\rm peak}_{e}(T)$ fit], and plotted as stars in Fig.~\ref{fig:taufitandcomp} (left).

\begin{table}[H]
\centering
\begin{tabular}{@{}ccccc@{}}
\toprule
\multirow{2}{*}{Prototype} & \multicolumn{2}{c}{Energy {[}eV{]}}  & \multicolumn{2}{c}{Cross section  {[}$10^{-15}$cm$^2${]}} \\
\cmidrule(lr){2-3} \cmidrule(lr){4-5}
          & Hist. max.     & $\tau^{\rm peak}_{e}(T)$ fit & Hist. max.                                       & $\tau^{\rm peak}_{e}(T)$ fit                           \\ \midrule
A         & $0.340\pm0.006$ & $0.341\pm0.003$       & $3.06\pm1.23$ & $3.49\pm0.92$ \\
B         & $0.302\pm0.008$ & $0.305\pm0.006$         & $0.66\pm0.29$ & $0.85\pm0.38$ \\
C         & $0.310\pm0.009$ & $0.313\pm0.007$         & $0.87\pm0.46$ & $0.98\pm0.51$ \\ \bottomrule
\end{tabular}
\caption{Energy and cross section associated to the largest population of traps in each of the Oscura prototype sensors computed as the maxima of the distributions in Fig.~\ref{fig:trapparams} [Hist. max.] and from the fits in Fig.~\ref{fig:taufitandcomp} (right) [$\tau^{\rm peak}_{e}(T)$].}\label{tab:trapparam}
\end{table}

As can be seen from Table~\ref{tab:trapparam}, the values of the trap parameters extracted from the primary peaks of the distributions in Fig.~\ref{fig:trapparams} [Hist. max.] and from the fits in Fig.~\ref{fig:taufitandcomp} [$\tau^{\rm peak}_{e}(T)$] are mutually consistent within errors. Furthermore, the parameters from prototype sensors B and C, both from the second fabrication batch, are also mutually consistent. This suggests that the kind of defects/contaminants inducing charge traps is related to the fabrication batch. It is worth noting that while the relative errors for energies are small, below 3\%, those for cross sections are significantly higher, ranging from 26\% in the best case to 53\% in the worst case.

The trap energies and cross sections reported in Table~\ref{tab:trapparam} are similar to those reported for hole traps associated to transition metals in p-type silicon~\cite{Claeys2018}, which are common materials used in semiconductor processing, for example: palladium (Pd), with $E_t=0.31$~eV and $\sigma=0.8\times10^{-15}$~cm$^2$, molybdenum (Mo), with $E_t=0.31$~eV and $\sigma=0.43\times10^{-15}$~cm$^2$, platinum (Pt), with $E_t=0.32$~eV and $\sigma=1\times10^{-15}$~cm$^2$, and silver (Ag), with $E_t=0.34$~eV and $\sigma=0.87\times10^{-15}$~cm$^2$. Although gettering techniques are implemented during the fabrication process to capture impurities, the use of the same equipment for productions involving transition metals could lead to unwanted metal contamination in the sensors.

\section{Effect of 1$e^-$ traps on DC measurements in skipper-CCDs}
Dark current (DC) is an irreducible exposure-dependent background for skipper-CCD detectors that originates from the thermal excitation of electrons from the valence band to the conduction band. As it constrains the lowest SER that can be achieved, estimating its value is important in applications where the science reach is limited by the one-electron background rate.

Single-electron traps within the skipper-CCD buried-channel constitute a source of SEEs, which can come from: 1) deferred charge from trap emission, see discussion in Section~\ref{sec:trapsignsinCCDs}, and 2) charge carriers generated through excitation processes that are enhanced by intermediate energy levels between the valence and conduction bands (midband states) associated to the traps~\cite{Shockley1952}. SEEs coming from deferred charge from trap emission are a background for DC measurements. However, SEEs from carriers generated through midband states contribute to the sensor's DC. The generation rate of the latter [carriers cm$^{-3}$ s$^{-1}$], in a fully-depleted CCD, can be expressed as~\cite{Janesick2001}
\begin{equation}
    U\sim \frac{\sigma v_{th} n_i N_t}{2\cosh{\frac{\vert E_t - E_i\vert}{k_BT}}}
\end{equation}
where $N_t$ is the concentration of traps at energy level $E_t$ [cm$^{-3}$], $E_i$ is the intrinsic (undoped) Fermi level [eV] and $n_i$ is the intrinsic carrier concentration [cm$^{-3}$]~\cite{Shockley1952}. $E_i$ and $n_i$ are computed as
\begin{equation}
    E_i=\frac{1}{2}\left[E_g+k_BT\ln{\left(\frac{N_v}{N_c}\right)}\right] \qquad {\rm and} \qquad n_i=\left[N_cN_v\exp{\left(-\frac{E_g}{k_BT}\right)}\right]^{1/2}
\end{equation}
assuming that the silicon band gap depends on temperature as $E_g(T)=1.1557-T^2[7.021\times 10^{-4}/(T+1108)]$~\cite{Varshni1967}. The temperature dependence of $N_{c(v)}$ is as in Eq.~\ref{eq:DOS} with $m^{h(e)}_{\rm dens}\simeq 0.94 (1.07)m_e$ for p-channel CCDs between 100K and 200K~\cite{green1990intrinsic}.

Using the energy and cross section associated to the largest population of traps found from the pocket-pumping measurements (Table~\ref{tab:trapparam}), we computed the contribution to DC from the single-electron traps obtaining $1.05\times10^{-14}\left(3.54\times10^{-10}\right)~e^-$/pix/day for 130K (150K); these numbers are several orders of magnitude below the expected DC, see discussion in Section~\ref{sec:DC}. Here, we assumed $N_t=2.15\left(n_{\rm traps}/V_{\rm bc}\right)$ with $n_{\rm traps}=8.5 \times 10^{4}$ the average number of traps in the buried-channel region of one sensor identified with the detection algorithm in the pocket-pumping measurements before applying the selection criteria, and $V_{\rm bc}=1.095\times 10^{-4}$~cm$^3$ the effective sensor's volume that was probed with the pocket-pumping technique; the factor 2.15 accounts for the traps in the second phase that were not probed and for a conservative 30\% dipole-detection inefficiency.

\subsection{DC measurements at surface and underground}\label{sec:DC}
A typical way to quantify dark current in skipper-CCDs is to acquire dark images with different exposure times, mask events within the images associated to any other source of background, compute the SER as a function of exposure time, and extract the slope, i.e. the dark single-electron rate, which represents an upper limit on the sensor's DC; see discussion in~\cite{sensei2022}. Performing these measurements underground allows us to minimize SEEs generated from external radiation interactions, which constitute a dominant background at the surface. In fact, the lowest single-electron rate ever achieved in a skipper-CCD is $1.6 \times10^{-4}~e^-$/pix/day~\cite{SENSEI:2020dpa}, reported by the SENSEI Collaboration from measurements in their setup at the MINOS cavern in the Fermi National Accelerator Laboratory (FNAL).

\begin{figure}[!ht]
    \centering
    \subfigure[]{\includegraphics[height=0.48\textwidth]{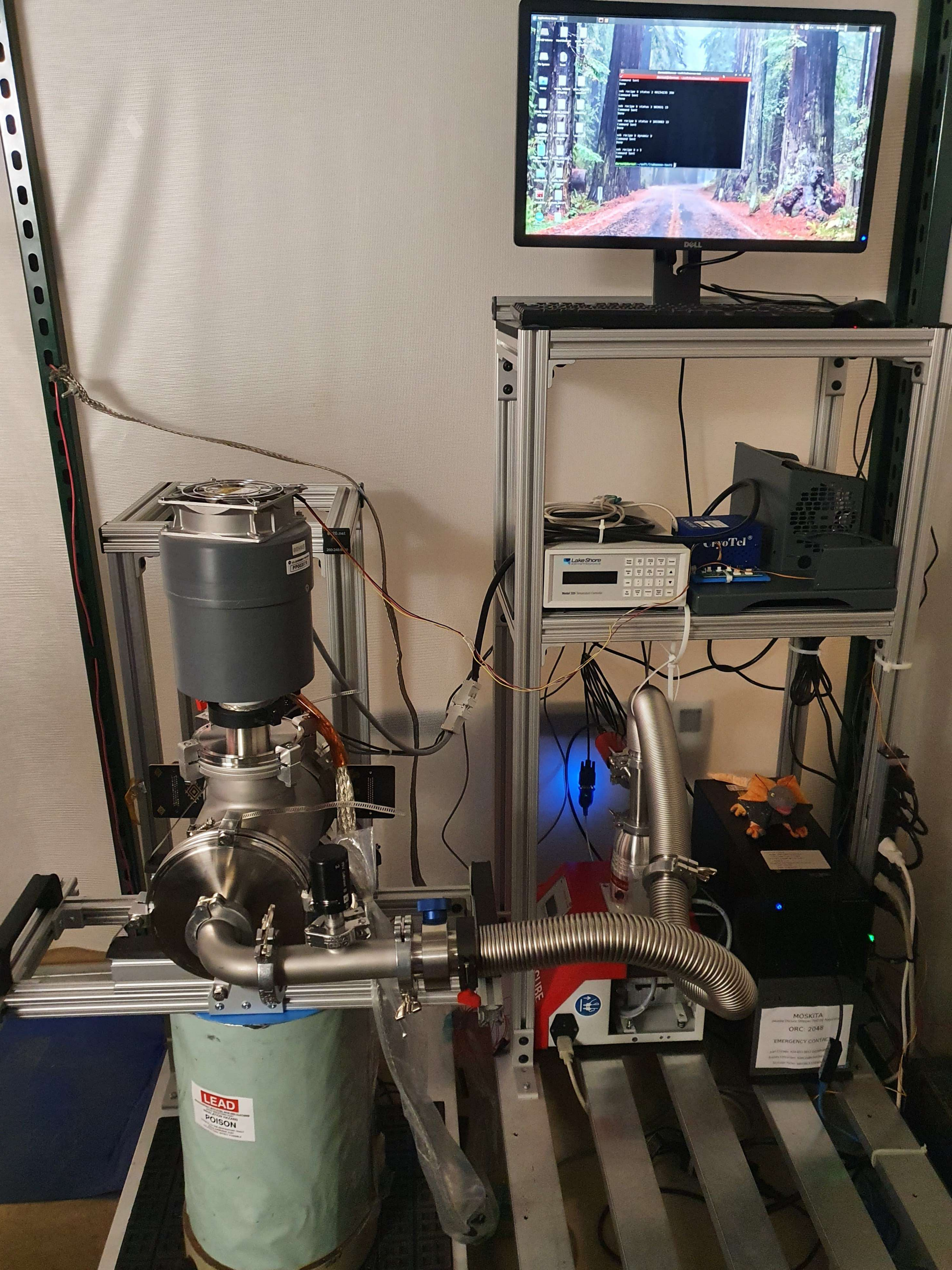}} \hspace{0.5cm}
    \subfigure[]{\includegraphics[height=0.51\textwidth]{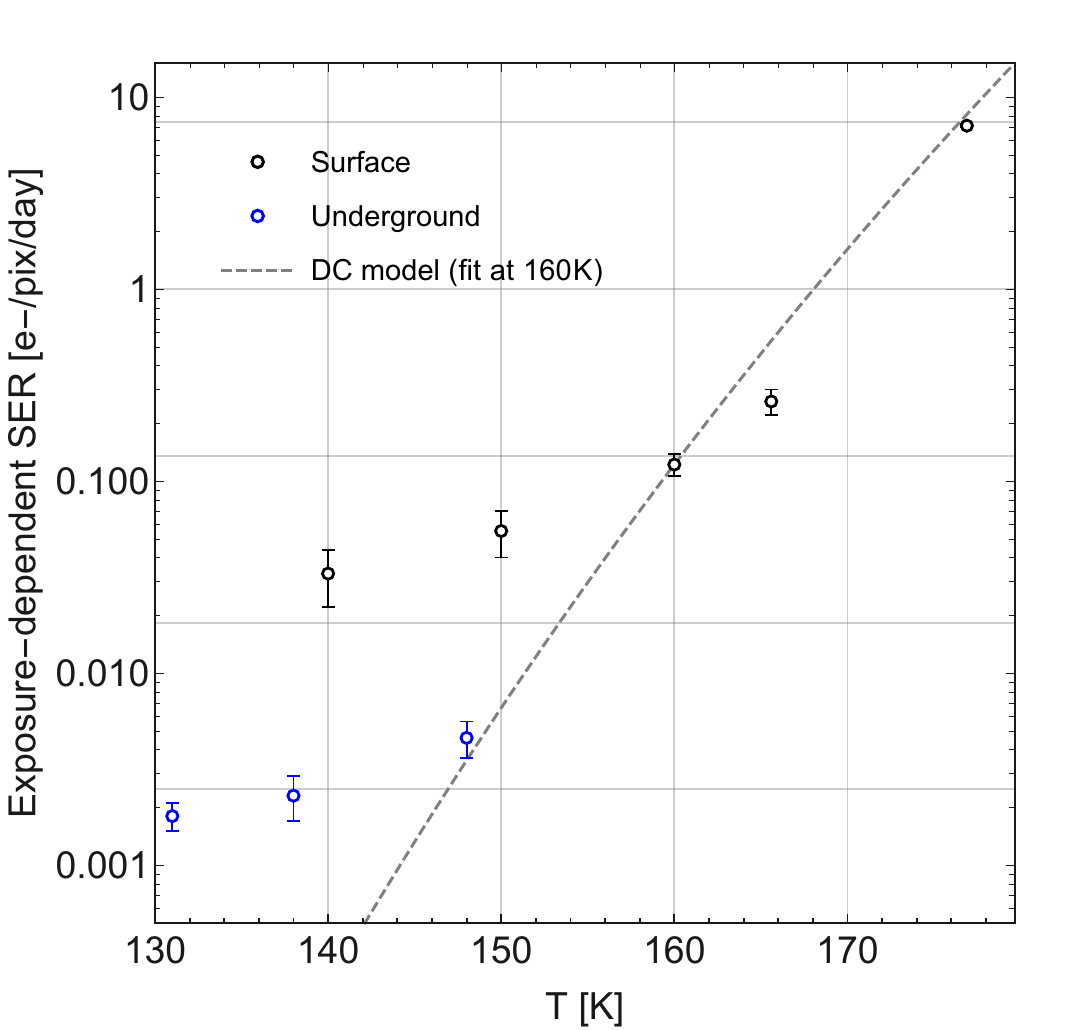}}
    \caption{Left: Dedicated setup with 2 inches of lead shield to perform DC measurements in the MINOS cavern at FNAL. Right: DC measurements from Oscura prototype sensors: at the surface for 140K, 150K and 160K~\cite{PM2023, OscuraSensors2023} (black) and at $\sim$100~m underground at MINOS for 131K, 138K and 148K (blue). For completeness, DC measurements at the surface with Oscura prototype A sensor in another testing setup for 166K and 177K (black) and the expected DC computed with Eq.~\ref{eq:expDC} and $D_{\rm FM}^{\rm 300K}=114$~pA/cm$^2$ (dashed) are shown.}\label{fig:dcmicrochip}
\end{figure}

In Refs.~\cite{PM2023, OscuraSensors2023} we presented DC measurements performed in a dedicated setup with 2 inches of lead shield at the surface with a Oscura prototype sensor from the same wafer as prototype A; these correspond to the circles in black at 140K, 150K and 160K in Fig.~\ref{fig:dcmicrochip} (right). The same setup was moved $\sim$100~m underground, to the MINOS cavern at FNAL; see Fig.~\ref{fig:dcmicrochip} (left). In that setup, we performed DC measurements with a Oscura prototype sensor from the same wafer as prototype C, following the previously discussed method. We acquired images varying the exposure time from 0 to 150~min, with 324~samples/pix. The exposure-dependent single-electron rates were computed from images acquired at 131K, 138K and 148K; these are shown in Fig.~\ref{fig:dcmicrochip} (right) as blue circles. The lowest value achieved was $(1.8\pm 0.3)\times10^{-3}~e^-$/pix/day at 131K.

The expected dark current in a CCD as a function of $T$ can be expressed as~\cite{Janesick2001}
\begin{equation}\label{eq:expDC}
    R_{\rm DC}(T) = \frac{A_{\rm pix} D_{\rm FM}^{T_0}}{q_e T_0^{3/2} e^{-E_g(T_0)/2k_BT_0}}~T^{3/2}e^{-E_g(T)/2k_BT} \times 86400~{\rm s/day}
\end{equation}
where $A_{\rm pix}$ is the pixel surface area [cm$^2$/pix], $q_e$ is the electron charge [C] and $D_{\rm FM}^{T_0}$ is the ``dark current figure of merit'' at $T_0$ [A/cm$^2$]. We fitted the measured DC at 160K with Eq.~\ref{eq:expDC} and found $D_{\rm FM}^{\rm 300K}=114$~pA/cm$^2$. The expected DC as a function of $T$ assuming this figure of merit is shown as a dashed line in Fig.~\ref{fig:dcmicrochip} (right); at 130K, the expected DC is $5.18\times10^{-6}~e^-$/pix/day, three orders of magnitude less than the measured DC with the Oscura prototype sensors.

\begin{figure}[!ht]
    \centering
    \subfigure[]{\includegraphics[width=0.49\textwidth]{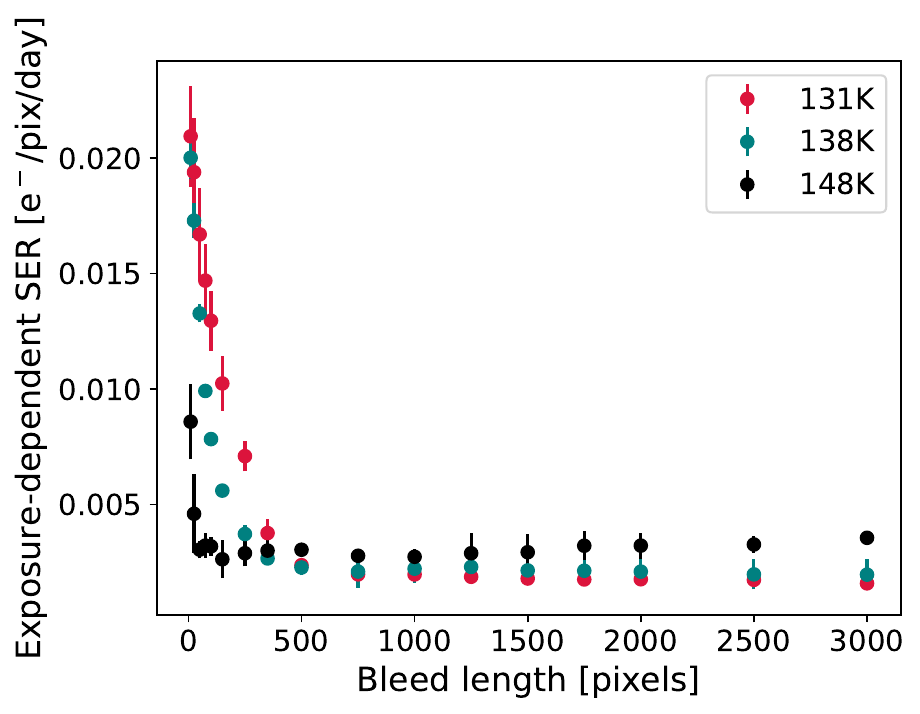}}
    \subfigure[]{\includegraphics[width=0.49\textwidth]{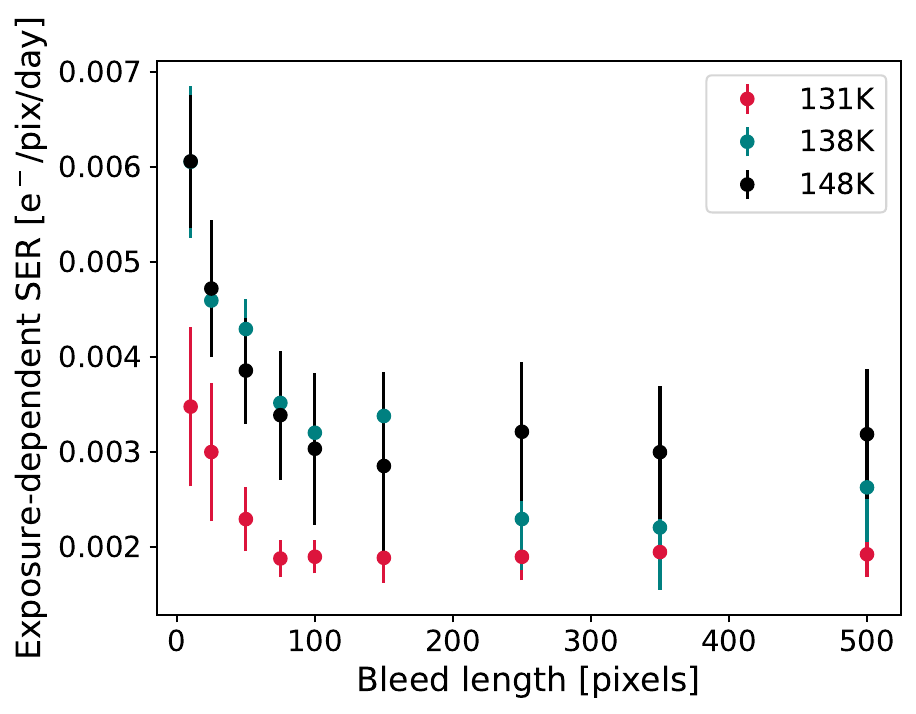}}
    \caption{Exposure-dependent single electron rate (SER) versus bleeding-mask length in the horizontal (left) and vertical (right) direction. In both cases, increasing the bleeding-mask length results in a decrease in the SER until it reaches a constant value. At lower temperatures, trap emission times are longer, resulting in a larger minimum bleeding-mask length for the SER to become constant; this is evident in the left plot.}\label{fig:compbleeds_temps}
\end{figure}

In the images taken underground with the Oscura prototype sensors, the SEEs originating from deferred charge from trap emission constitute a significant background for the DC measurements. To mitigate their impact, we implemented a ``bleeding zone'' mask for pixels upstream in the horizontal and vertical direction of any event with more than $20~e^-$, similar to what is done in skipper-CCD experiments searching for DM to discard events from charge-transfer inefficiencies~\cite{SENSEI:2020dpa, sensei2023}. To minimize the masked area of the images, we found the minimum bleeding-mask lengths in which the ``tails'' of deferred charge from trap emission did not impact the measured exposure-dependent SER. We measured this rate varying the horizontal (vertical) bleeding-mask length with a fixed vertical (horizontal) bleeding mask of 200 (1250) pixels, see Fig.~\ref{fig:compbleeds_temps}. The optimal mask length was determined as the minimum value after which the measured exposure-dependent SER becomes constant, being 1250 (250) pixels in the horizontal (vertical) direction.

\subsection{Monte Carlo simulations of deferred charge from trap emission}\label{sec:MCsims}
\begin{figure}[!ht]
    \centering
    \includegraphics[width=0.7\textwidth]{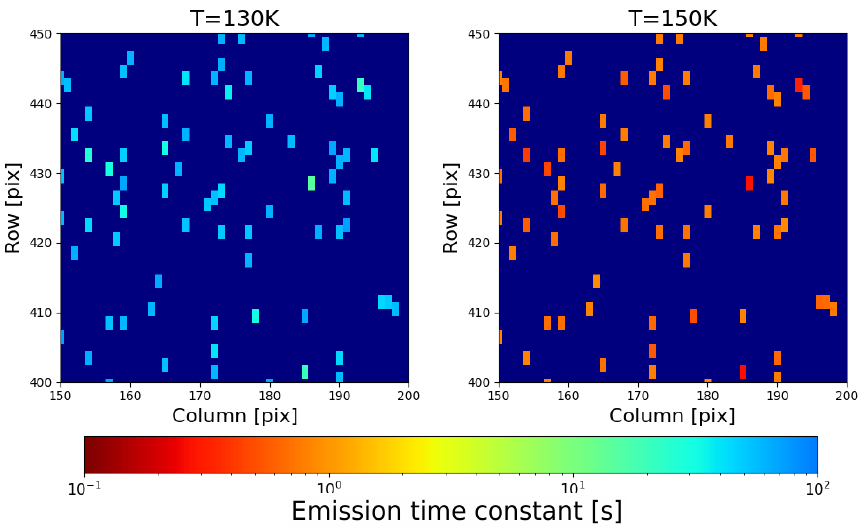}
    \caption{Region of the trap map of prototype sensor A for $T=130$K (left) and $T=150$K (right) used for simulating deferred charge from trap emission.}
    \label{fig:DC_impact_sim_result}
\end{figure}

We performed a Monte Carlo simulation to estimate the impact of deferred charge from single-electron trap emission on the measured exposure-dependent SER for the Oscura prototype sensors. For the simulation, we assumed that traps in the horizontal register have similar density, energy and cross section distributions as those in the vertical registers measured with the pocket-pumping technique, that the spatial distribution of traps is uniform and that sensors from the same fabrication batch have the same density and kind of traps. We generated trap maps at different temperatures with positions directly taken from the pocket-pumping measurements and emission-time constants computed using Eq.~\ref{eq:tautraps}, considering the trap parameters ($E_t$, $\sigma$) obtained from the measurements. Figure~\ref{fig:DC_impact_sim_result} shows a region of the detected trap maps from prototype sensor A measurements corresponding to two different temperatures.

The simulation is based on two sets of images taken with the Oscura prototype sensors: 1) underground, at 131K, with exposure times between 0 and 150~min, and 2) at surface, at 150K, with exposure times from 0 to 15~min. We simulated an ``underground'' (``at surface'') set of images, with each image containing the events with energy $\geq20e^-$ of the acquired image, a uniformly distributed exposure-independent SER of $1\times10^{-4}~(1\times10^{-2})~e^-$/pix and charge from a exposure-dependent SER of $1\times10^{-4}~(5\times10^{-2})~e^-$/pix/day, consistent with the exposure time of the acquired image.

\begin{figure}[!ht]
    \centering
    \includegraphics[width=1\textwidth]{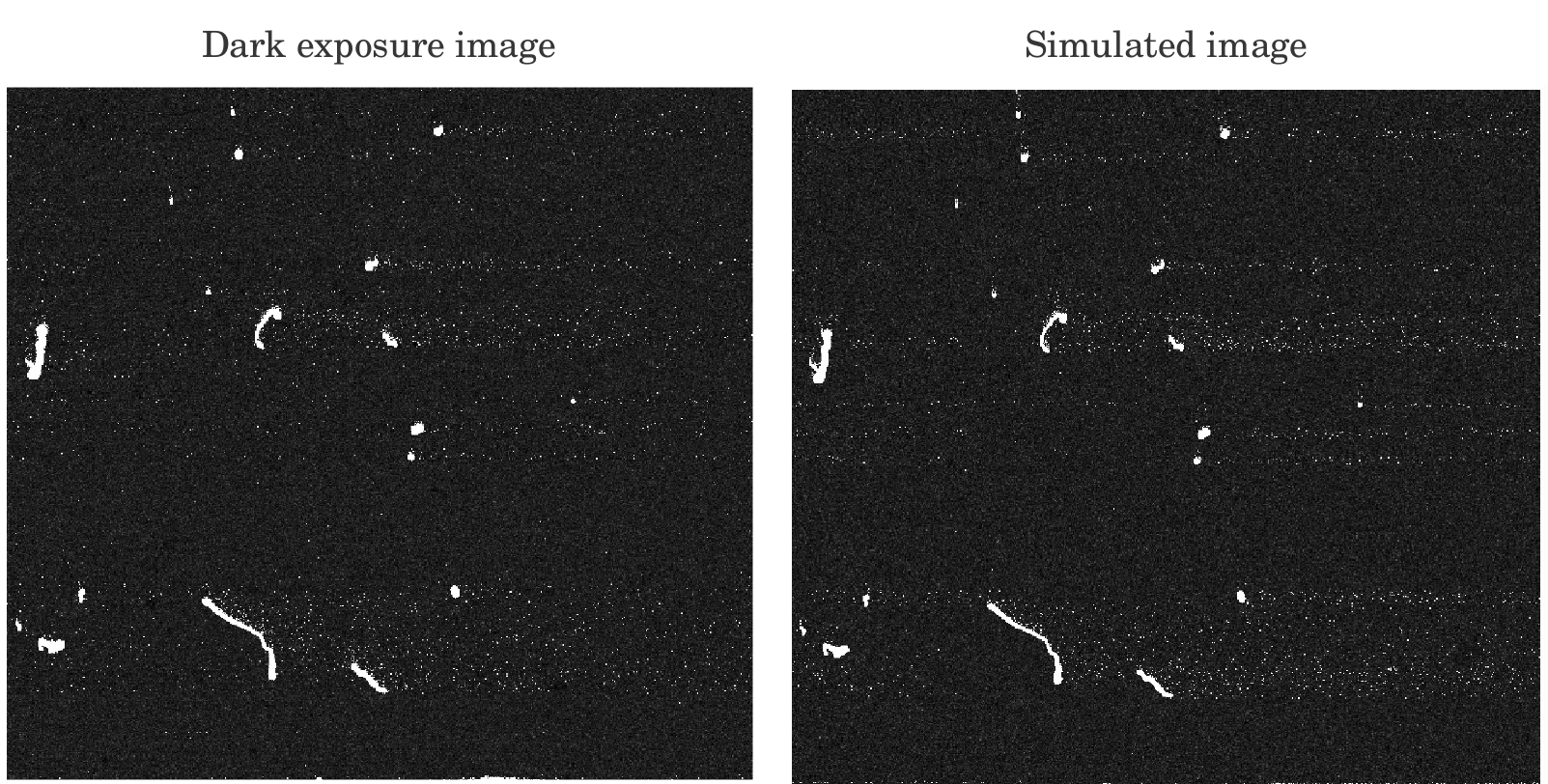}
    \caption{Comparison between a dark exposure image (left) and its corresponding simulated image (right). Deferred charge from multi-electron events can be observed in both images; however, in the right image, it was simulated using information from the detected trap map obtained through pocket-pumping measurements.}
    \label{fig:Comp_sim_deff_charge}
\end{figure}

Using these sets and the trap maps of prototype sensor C, we simulated two new sets accounting for the effects due to traps. For each event, we simulate the shifts of its constituting charge packets towards the readout amplifier. If the packet encounters a trap, a charge carrier is captured and released at a later time with a probability given by Eq.~\ref{eq:probtraps}. For simplicity, we assume the same capture probability for all traps and estimated the carrier density in its vicinity as in Ref.~\cite{Short2013}. In the simulation, carriers released from trap emission can be recaptured by subsequent traps and re-emitted at a later time. This causes a larger spread of carriers from trap emission within the image, which is more evident in the horizontal direction. In Fig.~\ref{fig:Comp_sim_deff_charge} we show a dark exposure image from one of the data sets and its corresponding image generated with the simulation.

We extracted the exposure-dependent SER on the simulated sets of images following the recipe outlined in Section~\ref{sec:DC}, using the optimal mask length. The extracted exposure-dependent SER in the simulated images without the effects of traps matches the simulated exposure-dependent SER of $1\times10^{-4}~(5\times10^{-2})~e^-$/pix/day for the “underground” (“at surface”) set. However, in the simulated images accounting for deferred charge from trap emission, we extracted a exposure-dependent SER of $(0.60\pm0.06)~e^{-}$/pix/day for the ``at surface'' simulated set and $(1.5\pm0.2) \times 10^{-3}~e^{-}$/pix/day for the ``underground'' simulated set. Both of these values are one order of magnitude larger than the simulated exposure-dependent SER.

These results show that, in sensors with traps that have emission times comparable to the readout time of consecutive pixels, depositions from trap emission can occur beyond the masked area, even with a conservative masking approach. Additionally, multi-electron events enhance trap capture. Overall, these factors can significantly impact the measured exposure-dependent SER. In fact, in the DC measurements at surface with Oscura prototype sensors, as presented in Refs.~\cite{PM2023, OscuraSensors2023}, the impact was minimized by increasing the image readout rate and selecting regions free of multi-electron events.

\section{Conclusions}\label{sec7}
We identified single-electron traps in the newly fabricated skipper-CCDs for Oscura. These traps have emission-time constants similar to the typical readout time of consecutive pixels, producing a ``tail'' of deferred charge observed in the images next to particle tracks. These ``tails'' consist mainly of single-electron depositions and can only be spatially resolved due to the sub-electron noise that can be reached with skipper-CCDs. Otherwise, deferred charge would only manifest as an increase in overall charge-transfer inefficiency and dark counts. In this sense, skipper-CCDs continue to provide insights into the understanding of dark-count sources.

We studied the buried-channel single-electron traps in three Oscura prototype sensors from two different fabrication batches and two different gettering methods, POCl$_3$ and ion implantation. The pocket-pumping technique was used to measure the position and emission-time constants of defects/contaminants associated to these traps at different temperatures. The trap characteristic parameters cross section and energy level were measured by fitting the temperature dependence of the emission times associated to each individual trap and to the primary peak of the $\tau_e$ distributions. Results from both analyses are consistent. The energy and cross section associated to the largest population of traps in each sensor are shown in Table~\ref{tab:trapparam}. These parameters are consistent within sensors from the same fabrication batch. Moreover, a secondary peak associated with a trap population with $\tau_e>0.1$s for $T>170$K is only observed in the sensor from the first fabrication batch. These results suggest that the type of defects/contaminants is more closely related to the fabrication process than to the implemented gettering.

The exposure-dependent SER was measured for a Oscura prototype sensor at the MINOS cavern at FNAL, yielding $(1.8\pm 0.3)\times10^{-3}~e^-$/pix/day at 131K. A procedure for finding the optimal bleeding-mask length to minimize the effect of charge traps encountered within the sensors was described. To estimate the impact of deferred charge from trap emission on exposure-dependent SER measurements, a Monte Carlo simulation of the trap capture and emission processes was implemented by using the trap parameters found from the pocket-pumping measurements. Results show that, even with a conservative masking approach, deferred charge from these traps can occur beyond the masked area and contribute to the measured exposure-dependent SER. More importantly, it provides an explanation to the rate measured underground of the Oscura prototype sensor. These results also suggest that the exposure-dependent SER of these sensors might be lower in lower background environments.

\appendix
\section{Dipole-detection algorithms}~\label{sec:dipdetection}
Algorithms designed to detect dipole signals against a flat background typically flag pixels with intensities that exceed or fall below a certain threshold established by the flat field signal. However, detecting dipoles becomes challenging when the dipole density increases or if the background is not flat. In this work, two different algorithms (A and B) were tested. Algorithm A subtracts the median of each row and computes the ``local'' standard deviation within a small window of pixels. The pixel intensity threshold is defined as a multiple of the local standard deviation. This code flags consecutive pixels if their absolute intensity is above the threshold and if one is positive and the other is negative. Algorithm B subtracts the median of each row and each column. It then asks for two consecutive pixels to be one positive and one negative, computes the dipole amplitude, scales it by a factor between 0 and 1 accounting for symmetry, and asks for the scaled amplitude to be above a certain threshold.

\begin{figure}[!ht]
    \centering
    \includegraphics{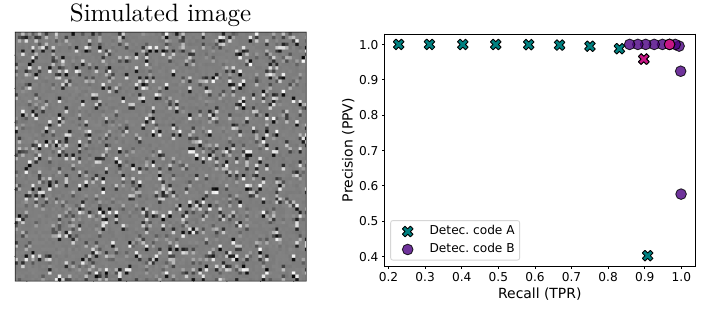}
    \caption{(Left) Simulated pocket pumped image showing a high density of dipoles. (Right) Precision-Recall curve for the two detection algorithms used in this work. The optimal detection threshold maximizes both metrics (red).}
    \label{comparison_codes}
\end{figure}

To select detection threshold values that yield the best dipole identification in an image with a high density of traps, a Monte Carlo simulation was made generating images with known numbers and positions of dipoles. By comparing the dipoles identified by the algorithms with the simulated ones, we computed the Precision and Recall detection metrics for each code and selected the detection threshold that maximizes their performance. Fig.~\ref{comparison_codes} shows one of the images with simulated dipoles (left) and the Precision-Recall curve for each algorithm when varying the detection threshold (right). The algorithm B, which accounts for the dipole symmetry, was found to have the better performance.

\acknowledgments
This work was supported by the resources of the Fermi National Accelerator Laboratory (FNAL), managed by Fermi Research Alliance, LLC (FRA), and the Lawrence Berkeley National Laboratory (LBNL), acting under Contract Nos. DE-AC02-07CH11359 and DE-AC02-05CH11231 with the U.S. Department of Energy, respectively.

\bibliographystyle{JHEP}
\bibliography{biblio.bib}

\providecommand{\href}[2]{#2}\begingroup\raggedright\begin{thebibliography}{10}

\bibitem{Janesick1990}
J.~{Janesick}, T.~{Elliot}, A.~{Dingizian}, R.~{Bredthauer}, C.~{Chandler}, J.~{Westphal} et~al., \emph{{New Advancements in Charge-Coupled Device Technology - Sub-Electron Noise and 4096 X 4096 Pixel CCDs}},  in \emph{CCDs in astronomy}, G.H.~{Jacoby}, ed., vol.~8, p.~18, Jan., 1990.

\bibitem{Tiffenberg2017}
J.~Tiffenberg, M.~Sofo-Haro, A.~Drlica-Wagner, R.~Essig, Y.~Guardincerri, S.~Holland et~al., \emph{{Single-Electron and Single-Photon Sensitivity with a Silicon Skipper CCD}}, \href{https://doi.org/10.1103/PhysRevLett.119.131802}{\emph{Phys. Rev. Lett.} {\bfseries 119} (2017) 131802}.

\bibitem{Rouven2016}
R.~Essig, M.~Fernández-Serra, J.~Mardon, A.~Soto, T.~Volansky and T.T.~Yu, \emph{{Direct detection of sub-GeV dark matter with semiconductor targets}}, \href{https://doi.org/10.1007/JHEP05(2016)046}{\emph{Journal of High Energy Physics} {\bfseries 2016} (2016) }.

\bibitem{Janesick2001}
J.~Janesick, \emph{{Scientific Charge-Coupled Devices}}, vol.~83 of \emph{Press Monograph}, SPIE (Jan., 2001), \href{https://doi.org/10.1117/3.374903}{10.1117/3.374903}.

\bibitem{sensei2022}
{\scshape SENSEI Collaboration} collaboration, \emph{{SENSEI: Characterization of Single-Electron Events Using a Skipper Charge-Coupled Device}}, \href{https://doi.org/10.1103/PhysRevApplied.17.014022}{\emph{Phys. Rev. Appl.} {\bfseries 17} (2022) 014022}.

\bibitem{Rouven2024}
P.~Du, D.~Egaña-Ugrinovic, R.~Essig and M.~Sholapurkar, \emph{Low-energy radiative backgrounds in ccd-based dark-matter detectors}, \href{https://doi.org/10.1007/JHEP01(2024)164}{\emph{Journal of High Energy Physics} {\bfseries 2024} (2024) }.

\bibitem{OscuraSensors2023}
B.A.~Cervantes-Vergara, S.~Perez, J.~Estrada, A.~Botti, C.R.~Chavez, F.~Chierchie et~al., \emph{{Skipper-CCD sensors for the Oscura experiment: requirements and preliminary tests}}, \href{https://doi.org/10.1088/1748-0221/18/08/P08016}{\emph{JINST} {\bfseries 18} (2023) P08016}.

\bibitem{Blouke1988}
M.M.~Blouke, F.H.~Yang, D.L.~Heidtmann and J.R.~Janesick, \emph{{Traps and Deferred Charge in CCDs}},  in \emph{Instrumentation for Ground-Based Optical Astronomy}, L.B.~Robinson, ed., pp.~462--485, Springer, 1988, \href{https://doi.org/10.1007/978-1-4612-3880-5_45}{DOI}.

\bibitem{Hall2014}
D.J.~Hall, N.J.~Murray, A.D.~Holland, J.~Gow, A.~Clarke and D.~Burt, \emph{{Determination of In Situ Trap Properties in CCDs Using a “Single-Trap Pumping” Technique}}, \href{https://doi.org/10.1109/TNS.2013.2295941}{\emph{IEEE Transactions on Nuclear Science} {\bfseries 61} (2014) 1826}.

\bibitem{Bilgi2019}
P.~Bilgi, \emph{{Optimization of CCD charge transfer for ground and space-based astronomy}}, Ph.D. thesis, California Institute of Technology, Apr., 2019.

\bibitem{Shockley1952}
W.~Shockley and W.T.~Read, \emph{{Statistics of the Recombinations of Holes and Electrons}}, \href{https://doi.org/10.1103/PhysRev.87.835}{\emph{Phys. Rev.} {\bfseries 87} (1952) 835}.

\bibitem{green1990intrinsic}
M.A.~Green, \emph{{Intrinsic concentration, effective densities of states, and effective mass in silicon}}, {\emph{Journal of Applied Physics} {\bfseries 67} (1990) 2944}.

\bibitem{HOLLAND1989}
S.~Holland, \emph{Fabrication of detectors and transistors on high-resistivity silicon}, \href{https://doi.org/https://doi.org/10.1016/0168-9002(89)90741-9}{\emph{Nuclear Instruments and Methods in Physics Research Section A: Accelerators, Spectrometers, Detectors and Associated Equipment} {\bfseries 275} (1989) 537}.

\bibitem{DALLABETTA1997}
G.~{Dalla Betta}, G.~Pignatel, G.~Verzellesi and M.~Boscardin, \emph{Si-pin x-ray detector technology}, \href{https://doi.org/https://doi.org/10.1016/S0168-9002(97)00612-8}{\emph{Nuclear Instruments and Methods in Physics Research Section A: Accelerators, Spectrometers, Detectors and Associated Equipment} {\bfseries 395} (1997) 344}.

\bibitem{Claeys2018}
C.~Claeys and E.~Simoen, \emph{Metal Impurities in Silicon- and Germanium-Based Technologies: Origin, Characterization, Control, and Device Impact}, Springer International Publishing (2018), \href{https://doi.org/10.1007/978-3-319-93925-4}{10.1007/978-3-319-93925-4}.

\bibitem{Varshni1967}
Y.~Varshni, \emph{Temperature dependence of the energy gap in semiconductors}, \href{https://doi.org/https://doi.org/10.1016/0031-8914(67)90062-6}{\emph{Physica} {\bfseries 34} (1967) 149}.

\bibitem{SENSEI:2020dpa}
{\scshape SENSEI} collaboration, \emph{{SENSEI: Direct-Detection Results on sub-GeV Dark Matter from a New Skipper-CCD}}, \href{https://doi.org/10.1103/PhysRevLett.125.171802}{\emph{Phys. Rev. Lett.} {\bfseries 125} (2020) 171802}.

\bibitem{PM2023}
B.~Cervantes-Vergara, S.~Perez, J.~D’Olivo, J.~Estrada, D.~Grimm, S.~Holland et~al., \emph{{Skipper-CCDs: Current applications and future}}, \href{https://doi.org/https://doi.org/10.1016/j.nima.2022.167681}{\emph{Nuclear Instruments and Methods in Physics Research Section A: Accelerators, Spectrometers, Detectors and Associated Equipment} {\bfseries 1046} (2023) 167681}.

\bibitem{sensei2023}
{\scshape SENSEI} collaboration, \emph{{SENSEI: First Direct-Detection Results on sub-GeV Dark Matter from SENSEI at SNOLAB}},  \href{https://arxiv.org/abs/2312.13342}{{\ttfamily 2312.13342}}.

\bibitem{Short2013}
A.~Short, C.~Crowley, J.H.J.~de~Bruijne and T.~Prod'homme, \emph{{An analytical model of radiation-induced Charge Transfer Inefficiency for CCD detectors}}, \href{https://doi.org/10.1093/mnras/stt114}{\emph{Monthly Notices of the Royal Astronomical Society} {\bfseries 430} (2013) 3078}.

\end{thebibliography}\endgroup

\end{document}